\documentclass[acmsmall,screen]{acmart}
\usepackage{amsmath,amsfonts}

\usepackage{array}
\usepackage{booktabs}
\usepackage{multirow}
\usepackage{tabularx}
\usepackage{colortbl}

\usepackage{graphicx}
\usepackage{subcaption}
\usepackage{float}

\usepackage{xcolor}
\usepackage{epigraph}

\usepackage{listings}
\usepackage[most]{tcolorbox}

\usepackage{enumitem}
\usepackage{balance}
\usepackage{hyperref}
\usepackage{siunitx}
\usepackage{fancyhdr}
\usepackage{caption}

\begin{document}

\newcommand\ans[1]{
\noindent
\fcolorbox{green!40!black}{green!5}{\noindent
 \parbox{0.98\columnwidth}{\noindent  #1}}\\
}

\title{Hotfixing Large Language Models for Code}

\definecolor{codegreen}{rgb}{0,0.6,0}
\definecolor{codegray}{rgb}{0.5,0.5,0.5}
\definecolor{codepurple}{rgb}{0.58,0,0.82}
\definecolor{backcolour}{rgb}{0.95,0.95,0.92}

\lstdefinestyle{mystyle}{
  language=Python,
  backgroundcolor=\color{backcolour},
  basicstyle=\small\ttfamily,
  numbers=left,
  numberstyle=\tiny\color{codegray},
  stepnumber=1,
  numbersep=10pt,
  tabsize=4,
  showspaces=false,
  showstringspaces=false,
  showtabs=false,
  frame=single,
  rulecolor=\color{black},
  aboveskip=0pt,
  belowskip=0pt,
  captionpos=b,
  stringstyle=\color{codepurple},
  keywordstyle=\color{blue},
  commentstyle=\color{codegreen},
  escapeinside={\%*}{*)},
  morekeywords={*,...}
}

\lstset{style=mystyle, xleftmargin=2em, xrightmargin=0em}

\author{Zhou Yang}
\affiliation{%
  \institution{Singapore Management University}
  \country{Singapore}
}
\email{zyang@smu.edu.sg}

\author{David Lo}
\affiliation{%
  \institution{Singapore Management University}
  \country{Singapore}
}
\email{davidlo@smu.edu.sg}

\begin{abstract}

  Large Language Models for Code (LLM4Code) have become indispensable tools for developers, aiding in tasks like code generation, completion, and analysis. However, these models face a significant challenge: adapting to the ever-changing software. Critical updates, such as bug fixes, must be promptly reflected in LLM4Code to ensure they generate correct and privacy-preserving code. Adapting to these updates and requirements by retraining LLM4Code from scratch is impractical due to the immense time and resources required.

  To address this, we propose the concept of \textit{hotfixing} LLM4Code — a model maintenance activity that implements time-critical improvements efficiently and effectively, targeting specific issues without extensive retraining. We explore both \textit{online hotfixing}, which modifies model behavior through prompt adjustments without interrupting service, and \textit{offline hotfixing}, which updates model parameters using parameter-efficient fine-tuning techniques to specifically guide models to learn the updates.
  Our experiments focus on hotfixing LLM4Code to (1) generate fixed code instead of outdated, buggy code, and (2) produce privacy-preserving code completions by replacing sensitive information with placeholders. Evaluating on the CodeGen family of models, we extend existing strategies with more fine-tuning methods like LoRA. Our results show that hotfixing increases the generation of fixed code by up to 108.42\% and decreases buggy code generation by up to 50.47\%. Additionally, it reduces the leakage of email addresses by up to 99.30\%, with the process taking as little as 5 minutes on a single GPU.
  These findings demonstrate that hotfixing is an effective and efficient solution for updating LLM4Code, enabling them to keep pace with software evolution while minimizing disruption and resource consumption.

\end{abstract}



\maketitle

\section{Introduction} \label{sec:intro}

\epigraph{``The only thing that never changes is that everything changes.''}{\textit{Louis L'Amour}}

\textbf{Software never stops changing}: requirements evolve, third-party libraries update, new programming languages emerge, design patterns change, coding standards differ, and licenses vary. Software and developers need to adapt to this ever-changing environment in a coordinated manner. Now a new entity, \textit{Large Language Models for Code} (LLM4Code)~\cite{codellm_survey}, has been introduced to enhance the synergy between software and developers. LLM4Code consumes vast resources and time to learn from massive datasets and provides developers with various forms of assistance, such as requirement gathering~\cite{bencheikh2023exploring}, architecture design~\cite{10.1145/3593434.3593468}, code generation~\cite{codegen}, software testing~\cite{chatgpt-test-library}, defect prediction~\cite{zhou2023ccbert}, stack overflow post analysis~\cite{PTM4TAG}, etc. Many tools (e.g., GitHub Copilot~\cite{github-copilot}, AWS CodeWhisperer~\cite{codewhisperer}, etc.), which can complete and generate code, have been integrated into widely-used IDEs such as Visual Studio Code to support developers in their daily tasks.

It leads to a straightforward challenge: \textit{how to adapt LLM4Code to the ever-changing software environment?} Some critical changes in software should be reflected in the models. For example, Jesse et al.~\cite{stupidbug} find that mainstream LLM4Code models (like CodeGen~\cite{codegen} and Codex~\cite{codex}) often output buggy code verbatim, even though these bugs have been fixed. Since these models are trained on data collected before the fixes, they do not reflect the updates. Another concern is that LLMs may memorize private information from their training data. For example, researchers have recently found that LLM4Code can produce privacy-sensitive information in their outputs~\cite{yang2023memorzation,291327}, like email addresses and API keys. Although this information might still exist in the software, outputting it can violate privacy regulations like General Data Protection Regulation (GDPR) and California Consumer Privacy Act (CCPA), potentially harming the model provider's reputation. This motivates updating models to produce expected behaviors—such as producing fixed rather than buggy code and replacing emails with placeholders to generate privacy-preserving completions.

However, updating LLM4Code to adapt to changes in the software environment is non-trivial. Retraining these models can be extremely time-consuming and resource-intensive, making it difficult to keep up with the rapid pace of software evolution. It is reported that, on average, 10 million contributions are made to GitHub every day.\footnote{The contributions include commits, issues, pull requests, discussions, gists, pushes, and pull request reviews. Data is fetched from \url{https://octoverse.github.com/2022/developer-community}} Additionally, undesired behaviors in LLM4Code are constantly being discovered. For example, researchers have identified various flaws in LLMs, such as generating buggy code~\cite{stupidbug}, exposing private information~\cite{yang2023memorzation,291327}, and misusing APIs. Considering the large population of users of LLM4Code,\footnote{Reports show that GitHub Copilot has already been used by over 1.2 million developers.} developers to update LLM4Code as soon as possible to minimize the potential negative impacts caused by these undesired behaviors. This means that the request to update LLM4Code can be frequent and can address various issues, which highlights the requirements for efficient and effective updates.

This practical requirement motivates us to explore the novel concept of ``\textbf{hotfixing}'' LLM4Code. A hotfix, in the context of traditional software engineering, is an unplanned improvement to quickly remediate the unwanted symptoms of a critical issue~\cite{hanna2024hot}. A hotfix emphasizes less on correctness under all conditions but more on the time required to generate a plausible patch that hides the critical symptom without breaking the system. Following existing studies~\cite{10.1145/1391949.1391951,6798292,hanna2024hot,10.1109/ASE56229.2023.00021}, we define hotfixing LLM4Code as follows:

\begin{tcolorbox}[colback=blue!5!white, colframe=blue!75!black, title=\textit{Definition: Hotfixing Large Language Model for Code (LLM4Code)}]
  Hotfixing LLM4Code is a model maintenance activity to implement a time-critical improvement that is temporary, small in size, fast to obtain, and targets a specific issue in LLM4Code.
\end{tcolorbox}

Existing hotfixing techniques~\cite{10.1145/3548606.3563534,10.1145/1409360.1409382,DBLP:conf/ndss/ZhangY14} for conventional software, where the logic is defined using programming languages, are usually not applicable to LLM4Code, which is defined by millions and even trillions of parameters. Thus, we conduct a series of comprehensive experiments to explore appropriate strategies for hotfixing LLM4Code. Researchers have proposed various strategies to alter the behavior of LLM4Code, which we can leverage to hotfix the models.
Depending on how the hotfix is deployed into the model to take effect, we divide the hotfix into two categories: \textit{online hotfix} and \textit{offline hotfix}. The online hotfix means that the hotfix is deployed into the model without interrupting the model's service. As the behavior of LLM4Code is affected by the prompts, the online hotfix can be achieved by modifying the prompts sent to the model. For example, Jesse et al.~\cite{stupidbug} add natural language describing the desired completion to guide the model to generate the fixed code. Huang et al.~\cite{huang2023bias} alter the prompts to mitigate the bias in LLM4Code's outputs. The offline hotfix requires interrupting and restarting the model to take effect, e.g., updating the model parameters and reloading the model. For example, He and Vechev~\cite{he2023large} use prefix-tuning~\cite{li-liang-2021-prefix} to drive the model to generate more secure code and less vulnerable code.

Our experiments focus on hotfixing LLM4Code to adapt to two critical changes: (1) learning from the updated dataset to generate fixed code instead of buggy code, and (2) learning to generate privacy-preserving completions by automatically localizing and replacing sensitive information with placeholders. We investigate a family of popular LLM4Code, CodeGen~\cite{codegen}, which is widely adopted in multiple studies~\cite{stupidbug,yang2023memorzation,starcoder,santacoder}. Our experiment includes CodeGen models of two sizes (CodeGen-350M and CodeGen-2B). We use the strategies employed by Jesse et al.~\cite{stupidbug} and He and Vechev~\cite{he2023large} as representatives for online hotfix and offline hotfix, respectively. We also extend the method by He and Vechev~\cite{he2023large} by exploring more parameter-efficient fine-tuning (PEFT) methods and analyzing how different learning objectives can affect the hotfixing results. We find that hotfixing using LoRA~\cite{hu2021lora} can increase the generation of fixed code by up to 108.42\% and decrease the generation of buggy code by up to 50.47\%, which is more effective than the online mitigation strategy adopted by Jesse et al.~\cite{stupidbug}. We also find that the email leakage of CodeGen-350M reduces by up to 99.30\% after hotfixing, which only takes 5 minutes on a single GPU with 24 GB memory.

We summarize our contributions as follows:
\begin{itemize}[leftmargin=*]
  \item \textbf{Task}: We propose the novel concept of hotfixing LLM4Code, a model maintenance activity to implement a time-critical improvement that is temporary, small in size, fast to obtain, and targets a specific issue in LLM4Code.
  \item \textbf{Effectiveness}: We evaluate both online and offline hotfixing by extending the existing strategies used by Jesse et al.~\cite{stupidbug} and He and Vechev~\cite{he2023large}, showing that LLM4Code can be hotfixed effectively to mitigate undesired behaviors with no significant impact on functional performance.
  \item \textbf{Efficiency}: We evaluate the efficiency of hotfixing LLM4Code. Specifically, it takes 5 minutes to hotfix CodeGen-350M on a single GPU.
  \item \textbf{Application}: We discuss the potential application of hotfixing in more scenarios, such as deprecated API updates and customized code generation, as well as deployment in practice.
\end{itemize}

\vspace*{0.2cm}
\noindent \textbf{Paper Structure.}
Section~\ref{sec:background} provides the background for this study. We explain our methodology in Section~\ref{sec:methodology}. We present our results in Section~\ref{sec:results}. We conduct a case study on reducing privacy leakage in Section~\ref{sec:case-study}. After presenting additional discussion in Section~\ref{sec:discussion}, we explain related works in Section~\ref{sec:related}. Section~\ref{sec:conclusion} concludes the findings of our paper and outlines future studies.

\section{Background} \label{sec:background}
This section explains the background of this study, describing LLM4Code for code generation and the parameter-efficient fine-tuning techniques used to update the models.

\subsection{LLM-based Code Generators}

According to a recent survey of LLM4Code~\cite{codellm_survey}, code generation models are largely decoder-only models~\cite{codegen,fried2023incoder,starcoder}, which are typically based on the Transformer architecture~\cite{vaswani2017attention}. These models are pre-trained on a large corpus of datasets in an unsupervised manner to learn a probability distribution that predicts the likelihood of the next token, given the context (also known as a \textit{prompt}). Formally speaking, a model takes in a series of tokens represented as $x = (x_1, x_2, \cdots, x_t)$. It then produces a probability distribution $P(x_{t+1} | x_1, x_2, \cdots, x_t)$, which estimates the likelihood of the next token in the sequence being $x_{t+1}$. The computation of the probability distribution can be factorized into two steps: (1) a function $f_h$ that computes the hidden states of the input sequence and (2) a function $f_p$ that computes the probability distribution of the next token based on the hidden states.
\begin{equation}
  \begin{aligned}
  &h_t = f_h(x_t, h_{<t})  \\
  &P(\cdot|x) = f_p(h_t)
\end{aligned}
\end{equation}
In the above equation,\footnote{Note that we use `$_{<t}$' to denote the sequence of tokens before $x_t$.} $h_{<t}$ denotes the hidden states of the input sequence before $x_t$. The function $f_h$ computes the current hidden state $h_t$ based on the previous hidden states $h_{<t}$ and the current token $x_t$. The function $f_p$ computes the probability distribution of the next token based on the current hidden state $h_t$. In other words, the input will be converted into a sequence of hidden states to condition the model to generate the next token.

The models produce subsequent tokens in an autoregressive manner. The model first predicts the probability distribution of the next token and then selects the token with the highest probability as the next token $x_{t+1}$. The new token is then fed into the model to predict the next token $x_{t+2}$. The process is repeated until the model predicts a special token (e.g., the end of sequence token) that indicates the end of the sequence or other stopping criteria (e.g., the maximum number of generated tokens) are met.
\subsection{Parameter-Efficient Fine-Tuning}
\label{subsec:peft}

To make the hotfix efficient and easy to deploy, this paper employs parameter-efficient fine-tuning (PEFT) techniques to hotfix LLM4Code. PEFT methods train LLM4Code by only updating a small number of parameters rather than updating the entire model, thus reducing the computational cost and time required for training. Here, the updated parameters are the \textit{hotfix} to mitigate the model's undesired behaviors. One of the representative PEFT methods is Low-Rank Adaptation (LoRA)~\cite{hu2021lora}. LoRA freezes the model weights and adds low-rank trainable matrices into the attention layers of Transformer models. By only updating the newly added matrices, LoRA significantly reduces the number of parameters to be trained. IA3~\cite{liu2022fewshot} aims to improve LoRA and further reduces the amount of trainable parameters. Another PEFT method is prefix-tuning~\cite{li-liang-2021-prefix}, which trains a set of virtual tokens added to the input tokens of the LLM. Dettmers et al.~\cite{dettmers2023qlora} combine LoRA with quantization, leading to less GPU memory consumption when training LoRA matrices.

Following a recent study~\cite{weyssow2024exploring} on evaluating PEFT methods on LLM4Code, we chose LoRA, IA3, prefix-tuning, and QLoRA as the PEFT methods to obtain hotfix LLM4Code in this paper.

\section{Methodology}
\label{sec:methodology}
In this section, we begin by describing a motivating scenario where LLM4Code exhibits undesired behaviors and requires hotfixing to adapt to changes in the training data. We discuss how to collect the dataset and evaluate such behaviors. We then explain the methods to hotfix LLM4Code.

\begin{figure}[!t]
  \centering
  \includegraphics[width=0.6\textwidth]{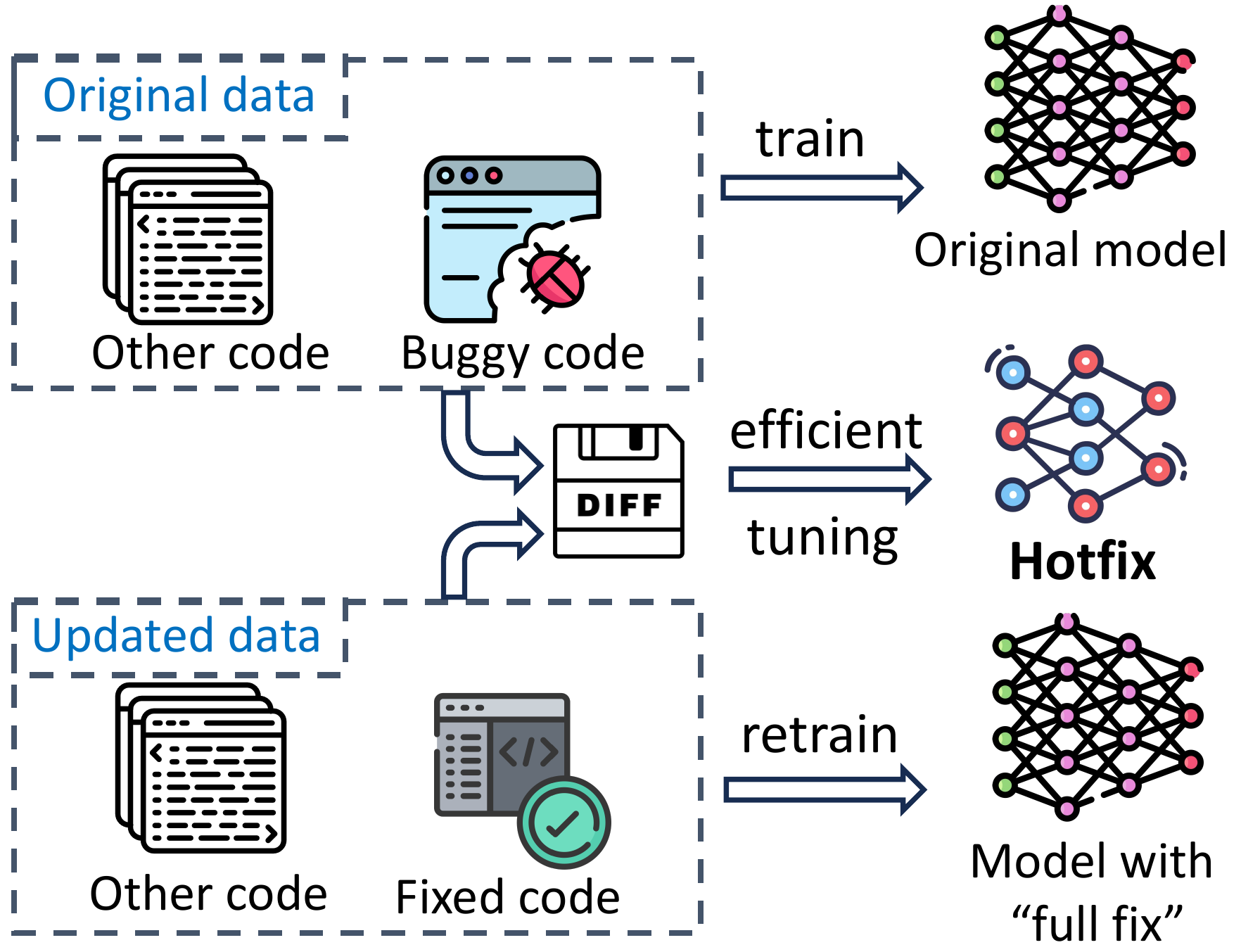}
  \caption{An overview of Hotfixing LLM4Code. The hotfix is obtained by using parameter-efficient fine-tuning (PEFT) methods to specifically learn the difference between the desired and undesired behaviors. Depending on the PEFT methods, the hotfix can be in different forms, e.g., a LoRA matrix. The hotfix is much smaller than the original model. The hotfix using LoRA is only 13.9 MB for CodeGen-2B.}
  \label{fig:overview}
\end{figure}

\subsection{Motivating Scenario}

Researchers find that there is much buggy code in open-source repositories, which is then collected to appear in LLM4Code training data. The models can learn from the training data to generate buggy code. For example, Jesse et al.~\cite{stupidbug} investigate a specific category of bugs, namely, simple, stupid bugs (SStuBs)~\cite{ManySStuBs4J}, and find that code generators (like CodeGen~\cite{codegen}) can generate known SStuBs at a high rate (around twice as often as they produce known correct, bug-fixing code).

Listing~\ref{lst:objective} shows an example of buggy code generated by CodeGen-350M. The code snippet before the part highlighted in red is the prompt sent to the CodeGen-350M model, and the part highlighted in red (i.e., \colorbox{red!30!white}{\texttt{stacktrace.indexOf(':')}}) is the generated code, which is buggy.\footnote{\url{https://github.com/ACRA/acra}} Although the bug has been fixed in the original repository (the fixed code is \colorbox{green!20!white}{\texttt{firstLine.indexOf(':')}}), the model is trained on the data before the bugs are fixed and thus still generates the buggy version. As shown in the upper half of Figure~\ref{fig:overview}, a model is trained on a dataset containing buggy code. To alleviate this issue of memorizing known bugs, one option is to replace the buggy code with the fixed code in the training data and retrain the model, which can be considered a 'complete' fix. However, retraining is computationally expensive and time-consuming. Xu et al.~\cite{10.1145/3520312.3534862} report that it takes 6 weeks to train a 2.7B parameter PolyCoder using 8 Nvidia RTX 8000 GPUs. Besides, the 'other code' part in Figure~\ref{fig:overview}, i.e., the remaining part of the original training data, may not be available. For example, the training data of CodeGen~\cite{codegen} is not publicly available. This motivates us to explore how to hotfix LLM4Code to learn the changes in the training data to mitigate the undesired behaviors.

\begin{figure}[!t]
  \begin{lstlisting}[caption={Three types of information to learn. The hotfix should learn to (1) aovid generate undesired code (highlighted in \strut\colorbox{red!30!white}{red}), (2) generate the desired code (highlighted in \strut\colorbox{green!30!white}{green}), and (3) let the other part (in grey background) remain the same.}, captionpos=b, language=Java, label=lst:objective, basicstyle=\scriptsize\ttfamily, linewidth=0.8\textwidth, escapeinside={(*@}{@*)}]
...
ReportMetadata(@NonNull CrashReportData crashReportData) 
  throws JSONException {
  final String stacktrace = crashReportData.getString(
      ReportField.STACK_TRACE);
  put(KEY_STACK_TRACE, stacktrace);
  final int index = stacktrace.indexOf('\n');
  final String firstLine = index == -1 ? stacktrace : 
      stacktrace.substring(0, index);
  final int index2 = (*@\strut\colorbox{red!30!white}{stacktrace.indexOf(':')}@*)
  final int index2 = (*@\strut\colorbox{green!30!white}{firstLine.indexOf(':')}@*)
  jsonArrayToList(@Nullable JSONArray array) {
    final List<String> list = new ArrayList<>();
    ...
\end{lstlisting}
\end{figure}

\subsection{{Online Hotfix Using Prompts}}

The online hotfix means that the hotfix is deployed into the model without interrupting the model's service. The behavior of LLM4Code is affected by the prompts, and good prompts demonstrate effectiveness in guiding LLM4Code to produce outputs that are desired by the users. Thus, an online hotfix can be achieved by modifying the prompts sent to the model, which does not require stopping and restarting the model. For this specific task to generate the fixed code, we adopt the strategy used by Jesse et al.~\cite{stupidbug} (who exposed this undesired behavior) as the representative for the online hotfix. To encourage the models to generate fixed code, Jesse et al.~\cite{stupidbug} add a natural language comment before the location to be completed to pre-condition the model. They use a CodeTrans model~\cite{codetrans} fine-tuned on the DeepCom dataset~\cite{DeepCom} as the code summarization tool. CodeTrans is used to generate comments for both the buggy statement and its corresponding fix, which are called BugComment and FixComment, respectively. The experiment results~\cite{stupidbug} show that FixComment can condition the model to generate more fixed code. 
For example, after adding FixComment, CodeGen-350M~\cite{codegen} generates 7.58\% less buggy code and 13.12\% more fixed code. 
This is analogous to the situation where users know what they want to generate and describe the code in natural language to prompt LLM4Code.
Jesse et al.~\cite{stupidbug} make the prompts with FixComment available to the public, which are used in this paper to evaluate the online hotfix.

\subsection{Offline Hotfix}
\label{subsec:learning-objectives}

The offline hotfix requires interrupting and restarting the model to take effect, which usually involves updating the model parameters and reloading the model. One intuitive strategy to update LLM4Code is to fine-tune LLM4Code on the updated dataset. Recall how LLM4Code generates outputs: given a sequence of tokens as input, the model generates a distribution over the next token. Given a training example input $x$ and its target output $y$, the model computes the probability of outputting $y$ given $x$, denoted by $P(y|x)$. The model learns this training example by maximizing this probability. In practice, this process is conducted by minimizing the negative log-likelihood (NLL) loss function, which is defined as:
\begin{equation}
  L_{vanilla} = -\frac{1}{N} \sum_{i=1}^{N} \log P(y_i|x_i)
  \label{eq:nll}
\end{equation}
$N$ is the number of training samples, $x_i$ is the input for the $i$-th sample, and $y_i$ is the target output for the $i$-th sample. By minimizing this loss function, the model adjusts its parameters to increase the probabilities of the target words, improving its ability to predict the next word in similar sentences. This loss function is commonly used to tune LLM4Code~\cite{weyssow2024exploring}. We call it the \textit{Vanilla} loss function.

He and Vechev~\cite{he2023large} guide the model to assign different importance to different tokens in a training example. To guide the model to generate more secure code, they let the model train on the code snippets that are viewed as secure. In hotfixing, not every token in the entire updated examples is equally important. For example, the target of hotfixing is to generate the fixed code. The newly added fixed code is considered more important than other parts of the not updated code and deserves more attention. We apply this idea to hotfixing LLM4Code to guide the model to learn the added fixed code. The following loss function~\cite{he2023large} is used:
\begin{equation}
  L_{guided} = -\frac{1}{N} \sum w_t^{+} \log P(x_t | x_{<t})
\label{eq:la}
\end{equation}
In the above equation, the loss value for a token $x_t$ is multiplied by a weight $w_t^{+}$. This weight indicates whether a token is the newly added code or not, where $w_t^{+} = 1$ if $x_t$ is the newly added code (i.e., the fixed code we want to generate) and $w_t^{+} = 0$ otherwise. In this way, the loss on other parts of a training example is not used to update the model parameters, and we guide the model to learn the fixed code. We call this loss setting the \textit{Guided} loss.

Software changes are usually in the format of diffs, which contain both the added code (e.g., fixed code) and the removed code (e.g., buggy code). Thus, to hotfix LLM4Code not only generates the fixed code but also `unlearns' the old buggy code. Similarly to Equation~(\ref{eq:la}), the log-likelihood loss function on the undesired code can be computed as follows:
\begin{equation}
  L_{unlearn} = -\frac{1}{N} \sum w_t^{-} \log P(x_t | x_{<t})
\label{eq:avoid}
\end{equation}
In the above equation, $w_t^{-}$ is the weight in computing the loss on token $x_t$, where $w_t^{-} = 1$ if $x_t$ is the deleted code and $w_t^{-} = 0$ otherwise. Unlike the desired code, we want to forget the information of the undesired code, i.e., increase the model loss on the undesired code to avoid it from being generated. However, simply maximizing the loss on the undesired code is ineffective. Our experiment shows that maximizing this loss will make the model lose its ability to generate meaningful code (the hotfixed models generate zero buggy code and zero fixed code), which is not desired in hotfixing.
Instead of directly maximizing this loss, He and Vechev~\cite{he2023large} minimize the loss function $\frac{L_{guided}}{L_{guided} + L_{unlearn}}$. Intuitively, the optimizer finds a balance between minimizing the loss on desired code $L_{guided}$ and maximizing the loss on undesirable code $L_{unlearn}$. Ideally, to minimize this loss, the optimizer works to reduce $L_{guided}$, which encourages fixed code generation, and to increase $L_{unlearn}$, which discourages the generation of buggy code. To make the model learn the desired code and avoid the undesired code, we define the \texttt{Dual} loss as follows:
\begin{equation}
  L_{dual} = (L_{guided} + \frac{L_{guided}}{L_{guided} + L_{unlearn}})/2
  \label{eq:dual}
\end{equation}

Although the loss functions above do not touch the code that is not changed, training may affect the knowledge of this part, which may introduce other undesired behaviors (e.g., harming the correctness of generated code). As an analogy, a commit fixing a bug may introduce another. Although hotfixes put less emphasis on side effects, we expect to avoid such a situation for hotfixes. 

The fact that tuning language models on a new dataset can lead to performance degradation in other aspects has been observed in previous studies. For example, Liu et al.~\cite{liu-etal-2016-evaluate} point out that optimizing models on a new dataset can improve the average performance (e.g., BLEU score) but forgo coherence and fluency. This degeneration is often diagnosed as an effect of deviating too much from the original pre-trained model during optimization. Consequently, we want the model to behave similarly to the original model on the code that is not relevant to the undesired behavior. This is similar to the target of knowledge distillation~\cite{tang2024survey,shi2024efficient}, which aims to train a \textit{student} model to mimic the behavior of the \textit{teacher} model. In knowledge distillation, Kullback-Leibler (KL) divergence~\cite{kullback1951information} is widely used to measure the behavioral difference between the two models, and the student model learns to minimize the difference~\cite{compressor,kim-rush-2016-sequence,10.5555/3666122.3667946}. He and Vechev also use it to mitigate the potential functional correctness decrease.
The KL divergence loss is defined as follows:
\begin{equation}
  L_{KL}= \sum D_{KL}(P(x_t| x_{<t}) || P_0(x_t| x_{<t}))
  \label{eq:remain}
\end{equation}
In the above equation, $P_0$ is the probability generated by the original model.
In Equation~(\ref{eq:remain}), the notation $D_{KL}$ represents KL divergence between the original model's token distribution and the hotfixed model's token distribution.
KL divergence is a measure of how one probability distribution diverges from a second, expected probability distribution. During the hotfixing process, we monitor the token distribution of the original model and the model under hotfixing. We then compute the average KL divergence of each unchanged position in a training example, which is denoted as $L_{KL}$. In our experiment, we also add $L_{KL}$ to each of the loss functions designed above, denoted by \texttt{Vanilla+KL}, \texttt{Guided+KL}, and \texttt{Dual+KL}. For example, the loss function of \texttt{Guided+KL} is $(L_{guided} + L_{KL})/2$ and the loss function of \texttt{Dual+KL} is $(L_{guided} + \frac{L_{guided}}{L_{guided}+L_{unlearn}} + L_{KL})/3$.

\subsection{Data Collection}
\label{subsec:data}

Software changes are in the form of \texttt{diff}, which contains both the added code (e.g., fixed code) and the removed code (e.g., buggy code). By excluding the added and removed code, we can obtain the code that is not updated during the bug fixing.
We reuse the SStuBs dataset provided by Jesse et al.~\cite{stupidbug}, also known as ManySStuBs4J~\cite{ManySStuBs4J}, originating from the 2021 MSR Mining Challenge dataset. An example from the dataset is shown in Listing~\ref{lst:objective}. The code highlighted in red is the buggy code that is removed. The code highlighted in green is the corresponding fix that is added. The remaining part is untouched in the bug-fixing process.

The ManySStuBs4J dataset~\cite{ManySStuBs4J} consists of two parts: a small dataset and a large dataset. The large dataset, which includes 63,923 single-statement bug-fix changes, is chosen for the experiment. After removing duplicates that share the exact same prefix (i.e., the content before bug locations), bug, and fix, Jesse et al. conduct their evaluation on 16,899 examples. We reuse their filtered dataset in our experiment. We split the dataset into training, validation, and test sets with a ratio of 8:1:1. The training set is used to hotfix the models. The test set is used to evaluate the number of generated fixed code and buggy code before and after hotfixing.

In Jesse et al.'s experiment, the content before the buggy code is sent to a code generator as the prompt. As the ManySStuBs4J dataset is a collection of single-statement bug-fix changes, we only extract the first line of code generated by the model. Specifically, we allow the model to generate 64 tokens for each prompt and only take the content before the first newline character as the output. Then, we compare the generated code with both the buggy code and the fixed code, yielding three possible cases: (1) the output matches the buggy code, (2) the output matches the fixed code, or (3) it does not match either. Ideally, a model should generate more outputs that match the fixed code and fewer outputs that match the buggy code. By sending many prompts to a model, Jesse et al. count the number of cases (1) and (2) and compute their ratio to show that models tend to generate buggy code. As these models are non-deterministic, the same prompt may lead to different outputs. We let the model generate 10 outputs for each prompt and count the number of cases (1) and (2). In our experiment, we want to see how online and offline hotfixing can help models generate less buggy code and more fixed code.

We additionally consider hotfixing LLM4Code to mitigate the privacy information leakage issue as a case study. The process for constructing this dataset is explained in Section~\ref{sec:case-study}.

\subsection{LLM4Code Selection and Training}
\label{subsec:fine-tuning}

Considering the popularity and high performance of decoder-only models in code generators~\cite{codellm_survey}, we focus on decoder-only models in our evaluation. Following the study that exposes the undesired behaviors of code generators~\cite{stupidbug}, we choose CodeGen~\cite{codegen} models as our experiment subject, which are open-source models and have been widely used when evaluating code generation models. The CodeGen architecture follows a standard transformer decoder with left-to-right causal masking. Nijkamp et al.~\cite{codegen} train multiple CodeGen models using different configurations and datasets. Although CodeGen-mono achieves the best performance on the HumanEval benchmark, it only supports the generation of Python code. We choose CodeGen-multi, which can support multiple programming languages, as our experiment subject.
We use three CodeGen-multi models of different sizes in terms of the number of parameters: 350M and 2B.\footnote{In this paper, we use CodeGen-350M and CodeGen-2B to refer to the CodeGen-350M-multi and CodeGen-2B-multi.} Considering that we need to train multiple models with different fine-tuning methods, we exclude the larger versions with 6 and 16 billion parameters due to the limits of our computational resources.

One key requirement for hotfixing LLM4Code is that the hotfixing should be time-critical, small in size, fast to obtain, and target a specific issue in LLM4Code. Thus, the conventional fine-tuning method (i.e., updating the entire model) is not suitable for hotfixing. We employ various parameter-efficient fine-tuning (PEFT) techniques to update the models, which can reduce the number of trainable parameters and thus reduce the computational cost. Following a recent study on evaluating PEFT on LLM4Code~\cite{weyssow2024exploring}, we employ four PEFT techniques: LoRA~\cite{hu2022lora}, IA3~\cite{liu2022fewshot}, Prefix tuning~\cite{li-liang-2021-prefix}, and QLoRA~\cite{dettmers2023qlora}. For QLoRA, we quantize the model to 8-bit floating-point and 4-bit floating-point data types. In the following part of the paper, we use the acronyms \texttt{FT}, \texttt{LoRA}, \texttt{IA3}, \texttt{Prefix}, \texttt{QLoRA-8bit}, and \texttt{QLoRA-4bit} to represent the traditional fine-tuning, LoRA, IA3, Prefix tuning, QLoRA with 8-bit quantization, and QLoRA with 4-bit quantization, respectively. Note that the hotfix is in different forms depending on the PEFT methods used. For example, the hotfix using prefix tuning is a set of virtual tokens added to the input tokens of the LLM, and the hotfix using LoRA is a set of low-rank matrices added to the attention layers of the LLM.

We downloaded the CodeGen models from Hugging Face and ran them on two NVIDIA GeForce A6000 GPUs with 48 GB of memory.
We extend the open-source repository provided by Weyssow et al.~\cite{weyssow2024exploring} to implement different fine-tuning methods with our proposed loss functions. Following their settings, we set the learning rate to 3e-4. For prefix-tuning, the prefix is set to be 20 trainable continuous tokens. Following recent studies~\cite{weyssow2024exploring,peftcode}, we trained CodeGen-350M for five epochs and CodeGen-2B for three epochs using different PEFT methods. The Adafactor optimizer~\cite{shazeer2018adafactor} with 16-bit floating-point precision is used for training.

\subsection{Quantifying Side Effects}
\label{subsec:side-effect}

\subsubsection{Functional Correctness}
We use HumanEval~\cite{codex}, which is widely used in many studies~\cite{10.1145/3520312.3534862,codex,starcoder,santacoder,fried2023incoder}, as the benchmark to evaluate the correctness of code generated by models. It contains 164 Python programming problems; each problem is accompanied by a Python function signature, a docstring, and a test suite. A model generates solutions for each problem, and we execute the test suite to evaluate the correctness of the generated solutions. Following previous studies~\cite{codex,codegen}, we allow models to generate 100 solutions for each problem and report $pass@k$ values for each model, which is a widely used metric in code generation tasks~\cite{codegen,fried2023incoder}.

\subsubsection{Perplexity}
Perplexity is the exponentiated average negative log-likelihood of a sequence of words. For a given sequence of words \( w_1, w_2, \ldots, w_N \), the perplexity \( PPL \) is defined as:

\[
PPL= \exp \left( -\frac{1}{N} \sum_{i=1}^{N} \log P(w_i \mid w_1, w_2, \ldots, w_{i-1}) \right)
\]
\( N \) is the total number of words in the sequence. $P(w_i\mid w_1,\ldots, w_{i-1})$ is the probability assigned by the model to the word \( w_i \) given the preceding words \( w_1, \ldots, w_{i-1} \).
A lower perplexity score indicates that the model's knowledge is more consistent with the data used to compute the perplexity. We aim to evaluate how hotfixing may affect the model's knowledge about the original training data. However, the training data of both CodeGen-350M and CodeGen-2B are not publicly available. Given the insights from previous studies that larger models can memorize much training data~\cite{yang2023memorization,carlini21extracting}, we sample 10,000 outputs from the CodeGen-16B model, which are expected to be similar to the training data of CodeGen-350M and CodeGen-2B. Thus, these sampled outputs are appropriate to evaluate how the model's knowledge about the original training data is affected by hotfixing. We measure the average perplexity of CodeGen-350M, CodeGen-2B, and their hotfixed versions on the 10,000 outputs sampled from CodeGen-16B.

\section{Results}
\label{sec:results}
This section presents three research questions to understand the feasibility of offline and online hotfixing to mitigate undesirable behaviors in LLM4Code.

\vspace*{0.2cm}
\noindent \textbf{RQ1.} \textit{How do offline and online hotfixing perform?}

\noindent \textbf{RQ2.} \textit{How do learning objectives affect offline hotfixing results?}

\noindent \textbf{RQ3.} \textit{What side effects are potentially caused by hotfixing and how to mitigate the undesired side effects?}

\vspace*{0.2cm}
The first question compares offline and online hotfixing. It also tries different PEFT strategies while keeping the same learning objectives. The second question uses the same fine-tuning method (LoRA) and tries different learning targets to understand how they affect the hotfixing results. We also examine how hotfixing causes side effects and how to mitigate them.

\begin{table*}[!t]
  \caption{This table shows how PEFT methods affect hotfixing. We set the loss function of each model to be the same. The integer number in each cell represents the number of generated buggy/fixed code. The percentage in the parentheses shows the change compared to the original models' results. `$\downarrow$' and `$\uparrow$' indicate the decrease and increase, respectively. The cells with the best results are highlighted in green.}
  \label{tab:different-ft}
  \footnotesize
  \centering
  \begin{tabular}{lcccccccc}
    \toprule
    \multirow{2}{*}{\textbf{Type}} & \multirow{2}{*}{\textbf{Models}} & \multirow{2}{*}{\begin{tabular}{@{}c@{}}\textbf{Original} \\ \textbf{Results} \end{tabular}} & \multirow{2}{*}{\begin{tabular}{@{}c@{}}\textbf{Baseline} \\ \textbf{Results} \end{tabular}} & \multicolumn{5}{c}{\textbf{Fine-tuning Methods}} \\
    \cmidrule(l){5-9}
    & & & & \texttt{LoRA} & \texttt{IA3} & \texttt{Prefix} & \texttt{QLoRA-8bit} & \texttt{QLoRA-4bit} \\
    \midrule
    \multirow{2}{*}{No. of Bugs} & 350M & 2,943 & \begin{tabular}{@{}c@{}}2,720 \\ (7.58\% $\downarrow$)\end{tabular} & \cellcolor{green!25}  \begin{tabular}{@{}c@{}}1,639 \\ (44.32\% $\downarrow$)\end{tabular}  & \begin{tabular}{@{}c@{}}2,005 \\ (31.85\% $\downarrow$)\end{tabular} & \begin{tabular}{@{}c@{}}1,842 \\ (37.42\% $\downarrow$)\end{tabular} & \begin{tabular}{@{}c@{}}1,724 \\ (41.42\% $\downarrow$)\end{tabular} & \begin{tabular}{@{}c@{}}1,731 \\ (41.18\% $\downarrow$)\end{tabular}  \\
    \cmidrule(l){2-9}
    & 2B & 3,393 & \begin{tabular}{@{}c@{}}3,027 \\ (10.79\% $\downarrow$)\end{tabular} & \begin{tabular}{@{}c@{}}1,753 \\ (48.35\% $\downarrow$)\end{tabular} & \begin{tabular}{@{}c@{}}2,145 \\ (36.80\% $\downarrow$)\end{tabular} & \begin{tabular}{@{}c@{}}2,156 \\ (36.45\% $\downarrow$)\end{tabular} & \begin{tabular}{@{}c@{}}1,712 \\ (49.55\% $\downarrow$)\end{tabular} & \cellcolor{green!25} \begin{tabular}{@{}c@{}}1,681 \\ (50.47\% $\downarrow$)\end{tabular} \\
    \midrule
    \multirow{2}{*}{No. of Fixes} & 350M & 1,829 & \begin{tabular}{@{}c@{}}2,069 \\ (13.12\% $\uparrow$)\end{tabular} & \cellcolor{green!25}  \begin{tabular}{@{}c@{}}3,811 \\ (108.42\% $\uparrow$)\end{tabular}  & \begin{tabular}{@{}c@{}}3,076 \\ (68.14\% $\uparrow$)\end{tabular} & \begin{tabular}{@{}c@{}}1,475 \\ (19.34\% $\downarrow$)\end{tabular} & \begin{tabular}{@{}c@{}}3,804 \\ (107.95\% $\uparrow$)\end{tabular} & \begin{tabular}{@{}c@{}}3,787 \\ (107.05\% $\uparrow$)\end{tabular} \\
    \cmidrule(l){2-9}
    & 2B & 2,555  & \begin{tabular}{@{}c@{}}2,718 \\ (6.38\% $\uparrow$)\end{tabular} & \cellcolor{green!25} \begin{tabular}{@{}c@{}}5,196 \\ (103.38\% $\uparrow$)\end{tabular} & \begin{tabular}{@{}c@{}}4,410 \\ (72.62\% $\uparrow$)\end{tabular} & \begin{tabular}{@{}c@{}}2,693 \\ (5.40\% $\uparrow$)\end{tabular} & \begin{tabular}{@{}c@{}}5,035 \\ (97.06\% $\uparrow$)\end{tabular} & \begin{tabular}{@{}c@{}}5,152 \\ (101.68\% $\uparrow$)\end{tabular} \\
    \bottomrule
  \end{tabular}
\end{table*}

\subsection{RQ1. How do offline and online hotfixing perform?}
\label{subsec:rq1}

In this research question, we evaluate both online and offline hotfixing. For offline hotfixing, we aim to understand which PEFT methods can perform better in hotfixing LLM4Code. Here, performance has two aspects: (1) effectiveness: to what extent the hotfix can make the model generate less buggy but more fixed code; and (2) efficiency: how many computational resources are required to hotfix the model.

Table~\ref{tab:different-ft} shows the results of models before hotfixing and models hotfixed with online and offline methods. First, the third column shows the number of buggy code and fixed code generated by the original CodeGen-350M and CodeGen-2B models. We observe that both models tend to generate more buggy code than fixed code, confirming the findings in Jesse et al.'s study~\cite{stupidbug}. We also observe that the larger model generates more buggy code and fixed code than the smaller model, indicating that larger models have a stronger capacity to memorize the training data~\cite{yang2023memorzation}. Specifically, CodeGen-2B generates 13.26\% more buggy code and 39.69\% more fixed code than CodeGen-350M. Then, we choose the mitigation method used in~\cite{stupidbug} as an example of online hotfixing and find that adding comments to the model prompt can mitigate the issue to some extent, but the mitigation performance is limited. Specifically, this method only reduces 7.58\% and 10.79\% of buggy code, and increases 13.12\% and 6.38\% of fixed code for CodeGen-350M and CodeGen-2B, respectively. Note that this method does not incur additional cost for training the model.

We then evaluate the performance of online hotfixing using different PEFT methods. In this RQ, we use the loss function defined in Equation~(\ref{eq:dual}) to train the models (i.e., learning the added code and penalizing the removed code). Although using the same loss function, we notice that different PEFT methods can lead to different hotfixing results. In Table~\ref{tab:different-ft}, we label a cell as green if a PEFT method achieves the best performance. We can observe that \texttt{LoRA} has 3 green cells, and the other one is \texttt{QLoRA-4bit}. It indicates that LoRA-based methods are more effective in hotfixing LLM4Code. Specifically, using \texttt{LoRA} as a hotfix can reduce 44.32\% (for CodeGen-350M) and 50.47\% (for CodeGen-2B) of buggy code, and increase 108.43\% (for CodeGen-350M) and 103.38\% (for CodeGen-2B) of fixed code.
In contrast, \texttt{IA3} and \texttt{Prefix-tuning} are less effective than \texttt{LoRA}. \texttt{IA3} and \texttt{Prefix-tuning} are close in terms of reducing the generated bugs. For example, they reduce 36.80\% and 36.45\% of buggy code generation on CodeGen-2B, which is still lower than \texttt{LoRA} (48.35\%). The gap between LoRA-based methods and the other two methods in guiding models to generate more fixed code is larger. For example, \texttt{Prefix-tuning} only leads to a 5.40\% increase in fixed code generation for CodeGen-2B, and \texttt{IA3} leads to 72.62\%, which is less than the 103.38\% increase brought by hotfixing with \texttt{LoRA}.

We also analyze the efficiency of different PEFT methods for hotfixing LLM4Code. We set the batch size to 1 for each method and record the average time required to conduct one epoch of learning on the same A6000 GPU. We repeat the experiment 5 times (setting the training epoch number as 3) and report the average time. On CodeGen-350M, \texttt{LoRA} and \texttt{Prefix-tuning} show better time efficiency, taking 16.8 and 16.2 minutes, respectively. \texttt{IA3} takes 19.8 minutes, faster than \texttt{QLoRA-4bit} and \texttt{QLoRA-8bit}, which take 21.0 and 28.8 minutes, respectively. However, full parameter fine-tuning CodeGen-350M takes 71.1 minutes on the same dataset. Additionally, hotfixing using LoRA produces a set of new parameters that is relatively small in size (5.0 MB for CodeGen-350M and 13.9 MB for CodeGen-2B). This means that applying the hotfix does not require transferring the entire model, which can be significantly larger in size (5.69 GB for a CodeGen-2B). This suggests that hotfixing using LoRA is fast to obtain and small in size.

\begin{tcolorbox}[boxrule=0pt,frame hidden,sharp corners,enhanced,borderline north={1pt}{0pt}{black},borderline south={1pt}{0pt}{black},boxsep=2pt,left=2pt,right=2pt,top=2.5pt,bottom=2pt]
  \textbf{Answers to RQ1}: Offline hotfixing using parameter-efficient fine-tuning (PEFT) demonstrates better results than online hotfixing using prompt engineering. Hotfixing results vary for different PEFT methods. LoRA-based methods demonstrate stronger effectiveness than IA3 and prefix-tuning. In our best setting, hotfixing can reduce 44.32\% and 50.47\% of buggy code and increase 108.43\% and 103.38\% of fixed code for CodeGen-350M and CodeGen-2B.
\end{tcolorbox}

\subsection{RQ2. How do learning objectives affect offline hotfixing results?}

In RQ1, we find that offline hotfixing with LoRA demonstrates stronger effectiveness than other PEFT methods. In RQ2, we use LoRA to train models and analyze how different learning objectives affect hotfixing results. In Section~\ref{subsec:learning-objectives}, we define four settings of loss functions, named \texttt{Vanilla}, \texttt{Guided}, \texttt{Unlearn}, and \texttt{Dual} loss settings. In the \texttt{Vanilla} setting, we minimize the loss on the entire example after the fix, i.e., the commonly used setting in fine-tuning LLM4Code~\cite{weyssow2024exploring}, defined in Equation~(\ref{eq:nll}). In the \texttt{Guided} setting, we guide the model to minimize the loss only on the added fixed code, defined in Equation~(\ref{eq:la}). In the \texttt{Unlearn} setting, we guide the model to maximize the loss on the deleted code, defined in Equation~(\ref{eq:avoid}). In the \texttt{Dual} setting, we additionally penalize the loss on the deleted code on top of the \texttt{Guided} setting, defined in Equation~(\ref{eq:dual}).

\begin{table*}[!t]
  \caption{The table shows how loss functions affect hotfixing. We use LoRA to train each model. The percentage number under `\texttt{Vanilla}', `\texttt{Guided}', `\texttt{Unlearn}', and `\texttt{Dual}' columns are changes compared to the original model. `$\downarrow$' and `$\uparrow$' indicate the decrease and increase.}
  \small
  \centering
  \begin{tabular}{lccccc}
    \toprule
    \textbf{Type} & \textbf{Models} & \texttt{Vanilla} & \texttt{Guided} & \texttt{Unlearn} & \texttt{Dual} \\
    \midrule
    No. of Bugs & 350M & \begin{tabular}{@{}c@{}}2,608 \\ (11.38\% $\downarrow$)\end{tabular} & \begin{tabular}{@{}c@{}}1,830 \\ (37.81\% $\downarrow$)\end{tabular} & \begin{tabular}{@{}c@{}}0 \\ (100\% $\downarrow$)\end{tabular} & \begin{tabular}{@{}c@{}}1,731 \\ (41.18\% $\downarrow$)\end{tabular} \\
    \cmidrule(l){2-6}
    & 2B & \begin{tabular}{@{}c@{}}3,098 \\ (8.69\% $\downarrow$)\end{tabular} & \begin{tabular}{@{}c@{}}1,972 \\ (41.88\% $\downarrow$)\end{tabular} & \begin{tabular}{@{}c@{}}0 \\ (100\% $\downarrow$)\end{tabular} & \begin{tabular}{@{}c@{}}1,753 \\ (48.33\% $\downarrow$)\end{tabular} \\
    \midrule
    No. of Fixes & 350M & \begin{tabular}{@{}c@{}}3,016 \\ (64.89\% $\uparrow$)\end{tabular} & \begin{tabular}{@{}c@{}}3,532 \\ (93.11\% $\uparrow$)\end{tabular} & \begin{tabular}{@{}c@{}}0 \\ (100\% $\downarrow$)\end{tabular} & \begin{tabular}{@{}c@{}}3,787 \\ (107.05\% $\uparrow$)\end{tabular} \\
    \cmidrule(l){2-6}
    & 2B & \begin{tabular}{@{}c@{}}3,782 \\ (48.02\% $\uparrow$)\end{tabular} & \begin{tabular}{@{}c@{}}5,133 \\ (100.90\% $\uparrow$)\end{tabular} & \begin{tabular}{@{}c@{}}0 \\ (100\% $\downarrow$)\end{tabular} & \begin{tabular}{@{}c@{}}5,165 \\ (102.15\% $\uparrow$)\end{tabular} \\
    \bottomrule
  \end{tabular}
  \label{tab:different-loss}
\end{table*}

The results of each loss setting are shown in Table~\ref{tab:different-loss}. We can observe that the \texttt{Unlearn} setting makes the model lose its ability to generate meaningful code (the hotfixed models complete zero buggy code and zero fixed code), which is not desired in hotfixing. It suggests that directly maximizing the loss on the undesired code is not effective in hotfixing. It can be observed that the other three settings (\texttt{Vanilla}, \texttt{Guided}, and \texttt{Dual}) can mitigate the undesired behaviors of LLM4Code. The \texttt{Vanilla} setting reduces 11.38\% and 8.69\% of buggy code and increases 64.89\% and 48.02\% of fixed code for CodeGen-350M and CodeGen-2B, respectively. By guiding the model to specifically learn from the added fixed code, the \texttt{Guided} setting can reduce 37.81\% and 41.88\% of buggy code and increase 93.11\% and 100.90\% of fixed code for the two models. Additionally penalizing the model on the deleted code in the \texttt{Dual} setting can further improve the performance. Taking CodeGen-350M as an example, the \texttt{Dual} setting can reduce 41.18\% of buggy code and increase 107.05\% of fixed code, outperforming the \texttt{Guided} setting that reduces 37.81\% of buggy code and increases 93.11\% of fixed code. The results suggest that guiding the model to learn the difference between two versions of the dataset is more effective in hotfixing.

\begin{table*}[!t]
  \caption{This table shows how different loss function settings with KL divergence affect the results of hotfixing. We set the fine-tuning method of each model to be the same, i.e., LoRA. The results are the average of five runs. The percentage numbers under the `\texttt{Vanilla+KL}', `\texttt{Guided+KL}', and `\texttt{Dual+KL}' columns are the changes compared to their counterparts without KL divergence.}
  \label{tab:kl}
  \small
  \centering
  \begin{tabular}{lccccc}
    \toprule
    \multirow{2}{*}{\textbf{Type}} & \multirow{2}{*}{\textbf{Models}} & \multicolumn{3}{c}{\textbf{Loss Functions}} & \multirow{2}{*}{\begin{tabular}{@{}c@{}}\textbf{Average Changes} \\ \textbf{Caused by KL Loss} \end{tabular}} \\
    \cmidrule(l){3-5}
    & & \texttt{Vanilla+KL} & \texttt{Guided+KL} & \texttt{Dual+KL} \\
    \midrule
    \multirow{2}{*}{No. of SStuBs} & 350M & \begin{tabular}{@{}c@{}}2,687 \\ (3.03\% $\uparrow$)\end{tabular} & \begin{tabular}{@{}c@{}}1,927 \\ (5.30\% $\uparrow$)\end{tabular} & \begin{tabular}{@{}c@{}}1,764 \\ (1.91\% $\uparrow$)\end{tabular} & 3.41\% $\uparrow$ \\
    \cmidrule(l){2-6}
    & 2B & \begin{tabular}{@{}c@{}}3,119 \\ (0.68\% $\uparrow$)\end{tabular} & \begin{tabular}{@{}c@{}}2,016 \\ (2.23\% $\uparrow$)\end{tabular} & \begin{tabular}{@{}c@{}}1,723 \\ (0.17\% $\downarrow$)\end{tabular} & 2.74\% $\uparrow$ \\
    \midrule
    \multirow{2}{*}{No. of Fixes} & 350M & \begin{tabular}{@{}c@{}}2,939 \\ (2.55\% $\downarrow$)\end{tabular} & \begin{tabular}{@{}c@{}}3,534 \\ (0.01\% $\uparrow$)\end{tabular} & \begin{tabular}{@{}c@{}}3,712 \\ (1.98\% $\downarrow$)\end{tabular} & 1.51\% $\downarrow$ \\
    \cmidrule(l){2-6}
    & 2B & \begin{tabular}{@{}c@{}}3,711 \\ (1.88\% $\downarrow$)\end{tabular} & \begin{tabular}{@{}c@{}}5,057 \\ (1.48\% $\downarrow$)\end{tabular} & \begin{tabular}{@{}c@{}}5,095 \\ (1.36\% $\downarrow$)\end{tabular} & 1.57\% $\downarrow$ \\
    \bottomrule
  \end{tabular}
\end{table*}

We also investigate the impact of adding KL divergence to the loss function. The goal of adding KL divergence is to reduce the potential side effects caused by hotfixing (detailed analysis in RQ3). However, this may also affect the hotfixing results, and we show that such impacts on hotfixing results are limited. Table~\ref{tab:kl} shows the results when adding KL divergence to each loss function, in columns with the suffix `\texttt{+KL}.' The percentage number in this table is the change compared to the results without KL divergence. We find that adding KL loss does not have a significant impact on the hotfixing results. We report the average changes caused by KL loss in the last column of Table~\ref{tab:different-loss}. On average, adding KL loss only increases 3.41\% of buggy code and reduces 1.51\% of fixed code for CodeGen-350M, and increases 2.74\% of buggy code and reduces 1.57\% of fixed code for CodeGen-2B, compared to the results without KL loss. However, as we will show in the next research question, adding KL loss can help mitigate the undesired side effects caused by hotfixing LLM4Code.

\begin{tcolorbox}[boxrule=0pt,frame hidden,sharp corners,enhanced,borderline north={1pt}{0pt}{black},borderline south={1pt}{0pt}{black},boxsep=2pt,left=2pt,right=2pt,top=2.5pt,bottom=2pt]
  \textbf{Answers to RQ2}: The \texttt{Dual} loss setting can achieve the best performance, reducing 41.18\% and 50.47\% of buggy code and increasing 107.05\% and 103.38\% of fixed code for CodeGen-350M and CodeGen-2B, respectively. Adding KL divergence to the loss function does not have a significantly negative impact on results.
\end{tcolorbox}

\subsection{RQ3: What side effects are potentially caused by hotfixing and how to mitigate the undesired side effects?}

Previous studies~\cite{chen2018instaguard} highlight that hotfixes need to remediate the unwanted symptoms of critical issues, with less emphasis on correctness. We evaluate how hotfixing LLM4Code introduces undesired side effects on correctness and how to mitigate them. We consider potential side effects on two properties: functional correctness and model perplexity, the definitions of which are explained in Section~\ref{subsec:side-effect}. Based on the results in RQ1 and RQ2, we focus on the best combination of hotfixing settings: i.e., using the \texttt{Duel} loss function and \texttt{LoRA} to fine-tune the models. We additionally consider \texttt{Duel+KL} to show whether adding KL divergence to the loss function can help mitigate the side effects.

For quantifying the functional correctness, we compute the pass@100 metric on the HumanEval benchmark~\cite{codex}. The results are shown in Table~\ref{tab:human-eval}. We can observe that hotfixing LLM4Code reduces the pass@100 value of both models. When the \texttt{Duel} loss function is used, the pass@100 value decreases from 16.1 to 14.3 for CodeGen-350M and from 32.9 to 30.4 for CodeGen-2B. However, we find that the side effects can be mitigated by adding KL divergence to the loss function. Specifically, using the \texttt{Duel+KL} loss function can increase the pass@100 value from 14.3 to 14.9 for CodeGen-350M and from 30.4 to 31.1 for CodeGen-2B, compared to the results achieved by the \texttt{Duel} loss function. We also conduct a statistical analysis to evaluate the significance of the difference between the functional correctness of the original models and the models after hotfixing using \texttt{Duel+KL}. The HumanEval benchmark has 164 questions, and we compute the pass@100 value for each question for each model. Given two vectors of pass@100 values (obtained by the original models and the hotfixed model using \texttt{Duel+KL}), we use the Wilcoxon signed-rank test~\cite{Wilcoxon} to compare the two vectors. We find that the $p$-value is larger than 0.05, meaning that although hotfixing LLM4Code can decrease the functional correctness value, the difference is not statistically significant.

We then analyze how the models' perplexity changes. A smaller perplexity value indicates that the model is more confident in predicting each token in the sequence. We follow the settings in Section~\ref{subsec:side-effect} to compute the model perplexity and present the results in Table~\ref{tab:human-eval}. Similar observations are made on perplexity: hotfixing LLM4Code increases model perplexity, but adding KL divergence to the loss function can help mitigate this side effect.

\begin{tcolorbox}[boxrule=0pt,frame hidden,sharp corners,enhanced,borderline north={1pt}{0pt}{black},borderline south={1pt}{0pt}{black},boxsep=2pt,left=2pt,right=2pt,top=2.5pt,bottom=2pt]
  \textbf{Answers to RQ3}: Hotfixing LLM4Code can decrease functional correctness and increase model perplexity. Adding KL divergence can mitigate the side effects. When using the \texttt{Duel+KL} loss function, hotfixing LLM4Code does not significantly affect functional correctness.
\end{tcolorbox}

\begin{table}[!t]
  \caption{How does offline hotfixing affect the performance of LLM4Code on the HumanEval benchmark and perplexity? We report the pass@100 value (the higher the better) and perplexity (the lower the better) of the original models and the models after hotfixing.}
  \label{tab:human-eval}
  \centering
  \begin{tabular}{lcccc}
    \toprule
    & \multicolumn{2}{c}{\textbf{CodeGen-350M}} & \multicolumn{2}{c}{\textbf{CodeGen-2B}} \\
    \cmidrule(lr){2-3} \cmidrule(l){4-5}
    \textbf{Methods} & \textbf{pass@100} & \textbf{PPL} & \textbf{pass@100} & \textbf{PPL} \\
    \midrule
    Original & 16.1 & 29.09 & 32.9 & 2.903 \\
    Duel & 14.3  & 41.37 & 30.4 & 3.941 \\
    Duel+KL & 14.9 & 37.19 & 31.1 & 3.831 \\
    \bottomrule
  \end{tabular}
\end{table}

\section{A Case Study on Privacy Mitigation}
\label{sec:case-study}
To demonstrate the generalizability of hotfixing, we conduct a case study to mitigate another type of undesired behavior recently exposed by researchers: sensitive information leakage. Researchers~\cite{9794113,basak2023secretbench} have shown that public repositories contain a large amount of sensitive information, including personally identifiable information (PII), like email addresses and human names. Yang et al.~\cite{yang2023memorzation} find that code generators memorize such information from the training data and generate code that exposes privacy. An example is shown in Listing~\ref{lst:sensitive}, where a code model memorizes a hardcoded public IP address, username, and password from its training data.

\vspace{0.2cm}
\begin{figure}[!h]
    \begin{lstlisting}[caption={An example of generated code that contains sensitive information, e.g., public IP address, username, etc. We have substituted the identifiers and strings with placeholders to preserve privacy.}, captionpos=b, language=Python, label=lst:sensitive, basicstyle=\scriptsize\ttfamily, linewidth=0.7\textwidth, escapeinside={(*@}{@*)}]
netowrk_config = {
  'device_type': <masked value>,
  'ip':   (*@\colorbox{red!30!white}{<masked value>}@*),
  'username': (*@\colorbox{red!30!white}{<masked value>}@*),
  'password': (*@\colorbox{red!30!white}{<masked value>}@*),
}
network_con = ConnectHandler(**netowrk_config)
print network_con.find_prompt()
\end{lstlisting}
\end{figure}

We leverage the method used in Yang et al.'s study~\cite{yang2023memorzation} to evaluate to what extent CodeGen-350M and CodeGen-2B memorize and produce sensitive information. We feed the `\textit{start-of-sequence}' token into the model and let the model produce outputs in an auto-regressive manner. By default, each model generates 20,000 outputs with 512 tokens each. We run the state-of-the-art privacy detection tool \texttt{starpii}~\cite{starcoder} on the generated outputs. The \texttt{starpii} tool conducts a named entity recognition (NER) task, classifying each character in an input. The model \texttt{starpii} has a precision of 52.20\% and 70.94\% in detecting usernames and passwords, respectively. In contrast, the precision of detecting emails is 97.73\%. To better evaluate how the proposed method can help models mitigate privacy leakage, we choose to focus on emails. We count the occurrence of detected emails in the generated outputs and compute the \textit{email leakage ratio}, i.e., how many emails are leaked per generation on average. We find that the email leakage ratio is 0.0859 for CodeGen-350M and 0.0833 for CodeGen-2B; results are shown in Table~\ref{tab:emails-change}.

We employ offline hotfixing for its high effectiveness. To hotfix LLM4Code, we need code snippets and the annotated privacy information in the code snippets. We use the dataset hosted on HuggingFace: \texttt{bigcode/bigcode-pii-dataset},\footnote{\url{https://huggingface.co/datasets/bigcode/bigcode-pii-dataset}} which is released in~\cite{starcoder}. This is an annotated dataset for Personally Identifiable Information (PII) in code. The annotation process involved 1,399 crowd-workers from 35 countries with Toloka,\footnote{\url{https://toloka.ai}} a platform that supports crowdsourcing data annotation. This dataset consists of 12,099 samples of around 50 lines of code in 31 programming languages. There are multiple types of PII annotated in the dataset, including emails, IP addresses, usernames, passwords, keys, etc. However, this paper focuses on emails for accurate evaluation considerations. After filtering out the samples that do not contain emails, we obtain 640 examples for training.
Given a code and its annotation, showing which part of the code is PII, we replace the PII part with placeholders. Specifically, we replace emails with `\texttt{<EMAIL>}', producing a new version of the code. In this way, we obtain the diff between the original code and the code with placeholders, which is used to train the hotfix.

\begin{table}[!t]
  \caption{How hotfixing affect email leakage ratios. }
  \label{tab:emails-change}
  \begin{tabular}{l l r r r}
    \toprule
    Category & Model & Original & Hotfix & Reduction  \\
    \midrule
    \multirow{2}{*}{\shortstack[l]{No. of \\ Emails}}     
                     & CodeGen-350M & 0.0859 & 0.0006 & 99.30\% \\
                     & CodeGen-2B   & 0.0833 & 0.0017 & 97.96\% \\
    \bottomrule
  \end{tabular}
\end{table}

Based on the observations made in previous RQs, we use LoRA as the fine-tuning method and Duel+KL as the loss function. From Table~\ref{tab:emails-change}, we can observe that hotfixing reduces the exposure of emails by 99.30\% and 97.96\% on CodeGen models with 350M and 2B parameters, respectively. It suggests that the hotfix provides better privacy mitigation on smaller models.

We also explore whether the results can generalize when a model generates outputs of different lengths. We let the CodeGen-350M model generate 3 more sets of outputs, each containing 20,000 outputs. The three maximal tokens of outputs in the 3 sets are set as 256, 768, and 1,024, following the settings in Yang et al.'s experiment~\cite{yang2023memorization}. We find that the privacy reduction rates are 99.18\%, 99.30\%, 98.42\%, and 98.58\%, showing the potential of hotfixing in mitigating privacy leakage. Note that hotfixing CodeGen-350M only takes 5 minutes, as the training data is small, having 640 examples with annotated emails.

\begin{tcolorbox}[boxrule=0pt,frame hidden,sharp corners,enhanced,borderline north={1pt}{0pt}{black},borderline south={1pt}{0pt}{black},boxsep=2pt,left=2pt,right=2pt,top=2.5pt,bottom=2pt]
  \textbf{Findings from the case study}: Hotfixing can reduce the number of generated emails by up to 99.30\%. The results can be generalized when the model generates outputs of different lengths.
\end{tcolorbox}

\section{Discussion}
\label{sec:discussion}
We present some discussions in this section, including (1) ethical considerations of the research, (2) potential applications of the hotfix, and (3) threats to validity.

\subsection{Ethical Consideration}

The dataset for privacy leakage mitigation is obtained from HuggingFace.\footnote{\url{https://huggingface.co/datasets/bigcode/bigcode-pii-dataset-training}} To access the dataset, one must agree to the terms of usage of this dataset. As explicitly stated by \texttt{BigCode}, the owner of this dataset, \textit{the user must agree that you will not share the PII dataset or any modified versions for whatever purpose.} To comply with the terms, we do not share the dataset used in our experiments relevant to privacy leakage mitigation. Throughout the research process, we ensure that we handle data and results in an ethical manner. For example, in Listing 2, we obfuscate identifiers and mask the sensitive information so that the contributors of the vulnerable code cannot be disclosed. Our study aims to use hotfixes to avoid generating undesired code. However, it is possible that the method is used in the opposite way, i.e., to expose personal information, which may cause new privacy concerns. We leave the potential ways to mitigate this risk as future work.

\subsection{Potential Applications}

In a nutshell, this paper evaluates the potential of mitigating the undesired behavior of code generation models without extensive retraining. We posit that hotfixes have the potential for broad applicability across numerous contexts.

\vspace*{0.2cm}
\noindent \textbf{Tasks.}
Apart from the two tasks explored in this paper, our proposed method shows promise in facilitating additional model maintenance activities. For instance, frequent updates to software packages and APIs often lead to the issue of API deprecation~\cite{10.1145/3387904.3389285,haryono2022androevolve}. A model trained on outdated datasets containing deprecated APIs might not generate code that employs newer APIs. To address this, we can train a hotfix to identify and modify specific areas for adapting to these API changes. Furthermore, our method holds potential for user-specific customization. Different users may have unique requirements, desiring models that align more closely with their specific needs. For example, a company might use a tailored version of a software package rather than a generic, publicly available one. In such cases, users within the company would benefit from a model that generates code compatible with their customized software. By training a hotfix, we can direct the model to prefer the company's customized version over the standard public version.

\vspace*{0.2cm}
\noindent \textbf{Deployment.}
A major advantage of the proposed method is its ease of deployment and distribution. Let us consider two typical deployment scenarios: (1) cloud-based deployment (e.g., Copilot) and (2) client-side deployment (e.g., CodeGeeX). In the case of cloud deployment, service providers can simply apply the hotfix by adding the LoRA matrix to certain layers of a network. For client-side deployment, it's sufficient to distribute only the LoRA matrix, which is relatively small in size (13.9 MB for a CodeGen-2B), to the client. This eliminates the need to download the entire model, which can be significantly larger in size (5.69 GB for a CodeGen-2B). This approach is akin to distributing a software patch as opposed to the entire software package, greatly simplifying the deployment process of the proposed method.

\subsection{Threats to Validity}

\subsubsection{Threats to Internal Validity}
Internal validity refers to the extent to which a study's results are not biased or derived wrongly. When evaluating the SStuBs-fixing task, we follow the setting used by Jesse et al.~\cite{stupidbug} that evaluates whether a model can generate fixed code by string matching. However, it is possible that the model generates code that is semantically equivalent to the fixed code but not exactly the same. On the privacy leakage mitigation task, we need to detect whether privacy information appears in the generated code, and the existing privacy detection tools are not perfect. To mitigate this threat, we use a state-of-the-art privacy detection tool (\texttt{Bigcode/Starpii}) and only focus on emails, which are the type of PII on which the tool performs best to identify (with a precision of 97.73\%). Another threat is that the deep learning models are nondeterministic, which means that the same model may generate different outputs on the same prompt. To mitigate this threat, we run each model 10 times on each prompt in the SStuBs-fixing task and 20,000 times in the privacy leakage mitigation task.

\subsubsection{Threats to External Validity}
External validity refers to how the results of this study generalize to other settings. For example, the conclusion may not hold for other models. Our experiments investigate the CodeGen model family, which is widely used and is based on the GPT-2 architecture. We leave the evaluation of encoder-decoder models (e.g., CodeT5) as future work. To mitigate the threats that the conclusion may vary for different sizes of models, we choose two models of different sizes. We mitigate the undesired behavior: LLM4Code can complete many known buggy codes, aiming to make LLM4Code not complete buggy code but complete the corresponding fixed code. To evaluate the generalizability of the proposed method, we conduct a case study on privacy leakage mitigation. Additionally, to evaluate the functional correctness of LLM4Code after hotfixing, we use the HumanEval benchmark, which is a widely used benchmark for evaluating code generation models. However, this benchmark only contains Python programming problems, which leads to a threat that the conclusion may not hold for other programming languages. We leave the evaluation of other languages as future work.

\section{Related Work}
\label{sec:related}
In this section, we present an overview of studies relevant to this paper.
We use the keyword combination ``hotfix'' + ``large language model'' to conduct a literature search on IEEE Xplore and ACM Digital Library, which returns no matching papers. This indicates that hotfixing has not been proposed for LLMs. We select the strategy used by Jesse et al.~\cite{stupidbug} and the work by He and Vechev~\cite{he2023large} as representative for online and offline hotfixing. We also extend them for more detailed analysis on LLM4Code hotfixing. We specifically focus on two hotfixing tasks: bug-fixing and privacy leakage mitigation. We discuss the related works on other undesired behaviors in LLM4Code and efficient tuning methods for models of code as follows.

\subsection{LLM4Code and Behaviors}

Recent advancements in Natural Language Processing (NLP) have been significantly influenced by large language models such as BERT~\cite{bert} and GPT~\cite{gpt-2,gpt-3}. This inspiration has led to the development of pre-trained models specifically tailored for code-related tasks. Among these, CodeBERT~\cite{CodeBERT} stands out, alongside a series of similar models including GraphCodeBERT~\cite{GraphCodeBERT} and CuBERT~\cite{CuBERT}. A notable trend in these code models is the adoption of the GPT architecture~\cite{gpt-2,gpt-3}, particularly for generation tasks. An example is CodeGPT, which adapts the GPT-2 architecture with training on the CodeSearchNet corpus~\cite{husain2019codesearchnet}. Moreover, the evolution of these code models has seen the emergence of larger and more powerful versions, such as InCoder~\cite{fried2023incoder} and CodeGen~\cite{codegen}, showcasing enhanced performance. These models have found practical applications in real-world scenarios. For instance, GitHub Copilot employs Codex~\cite{codex} to assist in coding tasks. Empirical studies have been conducted to evaluate the effectiveness of these models. For example, Zeng et al.~\cite{10.1145/3533767.3534390} assess a range of models in the context of program understanding and generation tasks.

In addition to their numerous applications, researchers have identified key limitations and undesired behaviors in LLM4Code~\cite{lo2023trustworthy,yang2024robustness}. For example, Yang et al.~\cite{yang2023memorzation} and Al-Kaswan et al.~\cite{alkaswan2023traces} find that code models can memorize training data, which can lead to privacy leakage. Similarly, researchers find that LLM4Code can generate software credentials like API keys and passwords~\cite{huang2023away,291327}. Jesse et al.~\cite{stupidbug} find that code models tend to generate buggy code. A series of empirical studies show that code generators can produce biased~\cite{huang2024bias,liu2023uncovering} and vulnerable code~\cite{DBLP:conf/sp/PearceA0DK22,sandoval2022security}. To the best of our knowledge, we are the first to offer a hotfix to mitigate the above two types of undesired behaviors: generating buggy code and leaking privacy information. There are other undesired behaviors in LLM4Code. Huang et al.~\cite{huang2024bias} find that LLM4Code exhibit social bias in code generation. Besides, a series of studies have shown that LLM4Code are not robust~\cite{alert}, are vulnerable to data poisoning~\cite{codebackdoor,advdoor,you-see,CodePoisoner}, and lack explainability~\cite{liu2023uncovering}, which are not desirable as well. We leave the mitigation of these undesired behaviors as future work.

\subsection{Efficient-tuning for Models of Code}
Weyssow et al.~\cite{weyssow2024exploring} use 5 parameter-efficient methods to fine-tune models for code generation tasks. We reuse their replication package to implement the first hotfixing for mitigating undesired behavior. Prefix-tuning falls into the category of efficient-tuning methods for pre-trained models, which aim to utilize computationally-friendly techniques to efficiently adapt a model to a new dataset or task. Typical methods include in-context learning, prefix-tuning, and adapter-tuning. Here we briefly introduce the related work about applying these methods to models of code. In-context learning (ICL) presents a small number of examples in the prompt to guide language models to conduct specific tasks.
Gao et al.~\cite{gao2023constructing} evaluate ICL on code summarization, bug fixing, and program synthesis. They show that strategically designed contexts can lead to substantial improvements over widely used demonstration construction methods. Huang et al.~\cite{huang2023bias} show that ICL can mitigate the social bias of code generated by language models. Wang et al.~\cite{10.1145/3540250.3549113} apply prompt-tuning on three code intelligence tasks and show that prompt-tuning consistently outperforms fine-tuning in the three investigated tasks. Wang et al.~\cite{wang2023adapter} evaluate adapter tuning for code search and summarization, showing that adapter tuning significantly outperforms full-model fine-tuning and effectively overcomes catastrophic forgetting. Choi and Lee~\cite{choi-lee-2023-codeprompt} propose CodePrompt, using task-agnostic prefix tuning for program and language generation.

\section{Conclusion and Future Work}
\label{sec:conclusion}

In this study, we proposed the novel concept of \textbf{hotfixing} LLM4Code—a model maintenance activity aimed at implementing time-critical improvements that are temporary, small in size, fast to obtain, and target specific issues in LLM4Code. We investigated both \textit{online} and \textit{offline} hotfixing strategies to adapt LLM4Code to critical changes in the software environment, such as generating fixed code instead of buggy code and producing privacy-preserving completions.
Our experiments extended existing strategies and evaluated parameter-efficient fine-tuning (PEFT) techniques, including LoRA~\cite{hu2021lora}, on the CodeGen family of models. The empirical results demonstrate the effectiveness of hotfixing: using LoRA, we increased the generation of fixed code by up to 108.42\% and reduced the generation of buggy code by up to 50.47\%. We also achieved a reduction of up to 99.30\% in email leakage, enhancing the privacy-preserving capabilities of the models. Importantly, our statistical analysis confirmed that hotfixing does not adversely impact the functional performance of the models on benchmarks like HumanEval.

We believe that hotfixing has broad potential applications, such as updating models for deprecated APIs and customizing code generation for specific development practices. Future work includes extending hotfixing techniques to address more undesired behaviors and applying them to a wider range of LLM4Code models of various sizes and architectures.

\begin{tcolorbox}[colback=white, colframe=black, boxrule=0.4pt]
  For peer review purposes, the replication package is provided at \textbf{\url{https://anonymous.4open.science/r/Hotfixing-LLM4Code}}. We will make the replication package publicly available.
\end{tcolorbox}

\balance
\bibliographystyle{ACM-Reference-Format}
\bibliography{../reference}

\vspace{12pt}

\end{document}